\documentclass[twocolumn,showpacs,preprintnumbers,prb,amsmath,amssymb]{revtex4}
\usepackage[dvips]{graphicx}

\newcommand{\bx}{$B_X$ }
\newcommand{\bxy}{$B_X$}
\newcommand{\rbb}{$R(B)$ }
\newcommand{\rbby}{$R(B)$}
\newcommand{\bite}{Bi$_2$Te$_3$/FeTe }

\newcommand{\bty}{Bi$_2$Te$_3$}

\newcommand{\bper}{$B^{\perp}$ }
\newcommand{\bpery}{$B^{\perp}$}
\newcommand{\bpar}{$B^{\parallel}$ }
\newcommand{\bpary}{$B^{\parallel}$}

\newcommand{\aby}{$\sim$}

\newcommand{\be}{\begin{equation} }

\newcommand{\bib}{\bibitem}

\newcommand{\ene}{\end{equation}}

\newcommand{\figr}{Fig.~\ref}

\newcommand{\hcu}{$B_{c2}$ }

\newcommand{\hcuy}{$B_{c2}$}

\newcommand{\sg}{superconducting }

\newcommand{\sscy}{superconductor}

\newcommand{\tc}{$T_{c}$ }

\newcommand{\tcy}{$T_{c}$}

\newcommand{\ybco}{Y$_{1}$Ba$_{2}$Cu$_{3}$O$_{7}$ }
\newcommand{\ybcoy}{Y$_{1}$Ba$_{2}$Cu$_{3}$O$_{7}$}
\newcommand{\ncco}{Ne$_{2-x}$Ce$_{x}$CuO$_{4}$ }
\newcommand{\nccoy}{Ne$_{2-x}$Ce$_{x}$CuO$_{4}$}

\begin{document}

\title{Anomalous oscillatory magnetoresistance in superconducting transitions}
\author{Milind N. Kunchur}
\email[Corresponding author email: ]{kunchur@sc.edu}
\homepage{http://www.physics.sc.edu/kunchur}
\author{Charles L. Dean} 
\affiliation{Department of Physics and Astronomy, University of South
Carolina, Columbia, SC 29208, U.S.A.}
\author{Boris I. Ivlev}
\affiliation{Instituto de F\'{\i}sica, Universidad Aut\'{o}noma de San Luis Potos\'{\i}, San Luis Potos\'{\i}, 78000 Mexico}

\date{\today}
\date{September 25, 2015} 

\begin{abstract}
We have discovered an oscillatory magnetoresistance phenomenon in 
a wide range of superconducting systems, with a 
periodicity that is essentially independent of temperature, transport current, 
magnetic field, and even material parameters. The nearly universal period points to a possible fundamental mechanism deeper than superconductivity itself, and may result from 
intrinsic pair-breaking mechanisms at sub-atomic length scales. 
\end{abstract}


\maketitle
The formation of the superconducting state involves a delicate congruence between the parameters of the electronic structure and the interactions between the charge carriers and excitations in the solid. Minute alterations in any of these conditions can sensitively alter the superconducting state as reflected by its transition temperature \tc  and its resistance $R$ in the dissipative regime. $R$ decreases when the order parameter strengthens and
provides a sensitive probe of small changes in the underlying electronic structure. 
Arising from flux vortex motion, fluctuations, and percolation between resistive and zero-resistance 
regions, $R$ is typically a monotonically rising function of 
$T$,  magnetic field $B$, and transport electric current $I$. In
particular, the magnetoresistance (MR) is usually positive. 

We discovered oscillations in MR 
of a type that has not been seen before, and with a universality across materials that suggests a possible novel electronic phenomenon at sub-atomic length scales. 
The effect was seen when \rbb was measured with very fine steps in $B$ under low-noise, low-excitation, and low-temperature-variation conditions. While initially observed in films of the electron-doped infinite-layer Sr$_{1-x}$La$_{x}$CuO$_{2}$ superconductor, 
we have since also observed these oscillations 
in another electron-doped cuprate \nccoy , in a hole doped cuprate \ybcoy , and even in the interface between a topological insulator (\bty) and a chalcogenide (FeTe). 
Despite the large range of material parameters, sample dimensions, $B$ orientations, and 
a temperature range covering 4--74 K,  the period remains essentially the same (0.11--0.15 T). 
Not all samples tested show the effect, and some samples have a very weak oscillation amplitude. We suggest a possible 
connection between the observed phenomenon and mutual cancellation of subatomic-level pair breaking mechanisms. 

The main measurements in this study are on c-axis-oriented epitaxial thin films of
Sr$_{0.88}$La$_{0.12}$CuO$_{2}$ (SLCO) 
deposited on heated KTaO$_3$ substrates by rf magnetron sputtering 
followed by an oxygen reduction step. 
SLCO sample A had the parameters: 
thickness $d$=31 nm, width $w$=13.6 $\mu$m, length $l$=512 $\mu$m,
midpoint transition temperature \tcy = 23.3, and a transition width (10--90\% of normal resistance) of $\Delta T_c \approx 2.5$ K. 
Sample B had: $t = 61$ nm, $w=4 \ \mu$m, $l=100\  \mu$m, \tcy =  26.5 K, and  $\Delta T_c \approx 2.5$ K. 
The other sample included in this study is a c-axis-oriented epitaxial thin film of \ybco 
(YBCO) deposited on a SrTiO$_3$ substrate by the pulsed-laser-deposition process, 
with $t = 50$ nm, $w=4 \ \mu$m, $l=70\  \mu$m, \tcy=78.6 K and $\Delta T_c \approx 9$ K.
All samples were four-probe bridges patterned by projection photolithography and argon-ion milling. All samples were characterized by broad transitions. 

The cryostat was a Cryomech PT405 pulsed-tube closed-cycle cryocooler, fitted with a 1.2 Tesla GMW 3475-50 water-cooled copper electromagnet.
Regenerator materials used in a cryocooler can have a slightly $B$ dependent heat capacity, which can cause the cooling power, and potentially $T$, to vary with $B$ (although never in an oscillatory manner). In our system, the cold head is far removed from the magnet; a 22 cm long copper rod protrudes from the second-stage heat station into the magnet poles. The sample,  along with calibrated cernox and diode thermometers, are mounted in close proximity at the end of this copper rod (a diode sensor on the second-stage heat station serves as a tertiary indicator). A hall sensor measured $B$ and the current supplied to the electromagnet serves as a secondary indicator of $B$. 
The standard active temperature controller (which may produce $T$ oscillations and add electrical noise) was disconnected. Instead $T$ was set with 
a constant stable heater voltage;  
long-term temperature drifts were addressed by infrequently applying 
small precalculated corrections to this heater voltage, when $T$ drifted by more than 5 mK. 

A cryocooler's cooling power oscillates with the compressor cycle. 
To overcome this, all measurements were conducted synchronously at a fixed phase point 
of the compressor cycle. This was achieved by housing the compressor in an acoustically isolated enclosure with a microphone inside to detect the repetitive sound. This signal (after preamplification, active rectification, and pulse-shaping) triggered a pulse generator (with adjustable height and delay), which in turn triggered all other instruments (voltmeters, oscilloscopes, other pulse generators, etc.).
The reliability of this system's measurement of $R$, and 
stability of $T$ over time and against changes in $B$ was extensively checked with 
resistors as ``test samples'' and through the continuous monitoring of all three thermometers. 

All \rbb data represent 
four-probe current-direction-reversed resistance measurements at a constant dc $I=12.8 \mu$A.
All data are completely reversible with respect to changes in $I$, $T$, and $B$.
All data also lie in the Ohmic response regime (except for the one set of variable-$I$ curves). 
To minimize electrical noise, the dc current source consisted simply of an alkaline battery and 
a large series resistor, serving as a ballast to hold the current constant. 
(A Hewlett Packard 5532A dc power supply, 
feeding through an RC low-pass filter, replaced the battery for the one set of variable-$I$ curves.)
Except where noted, the sample voltage was measured with a Keithley model 2182A nanovoltmeter and other voltages measured with Keithley model 2000 multimeters, with each quantity averaged over \aby 30 readings (individual readings had integration times of 17 ms). The single confirmatory \rbb curve measured with pulsed signals 
utilized a Wavetek model 801 pulse generator, in-house built electronics, 
and a LeCroy LT 322 digital storage oscilloscope.

\begin{figure}[ht]
\includegraphics[width=0.95\hsize]{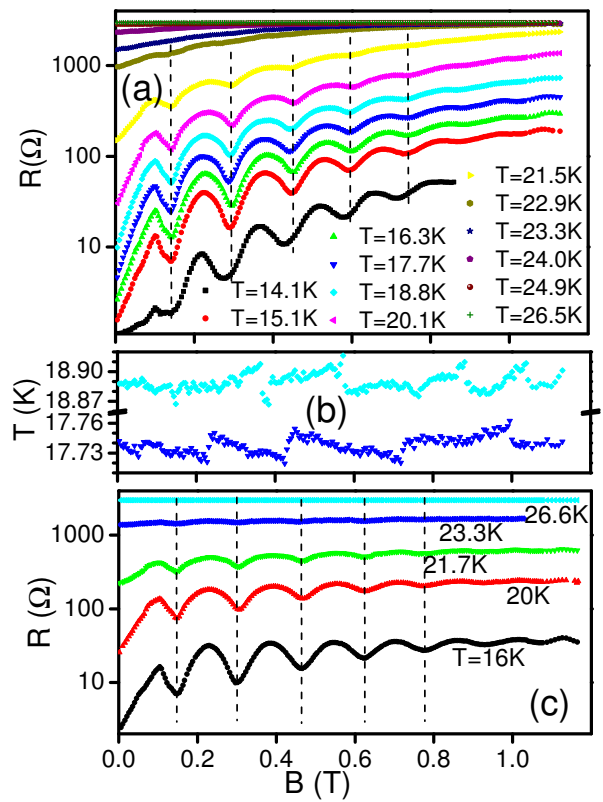} 
\caption{\em (a) \bper magnetoresistance curves in SLCO sample A for various temperatures.
Vertical dashed lines pass through the minima  $B_X$. 
(b) Temperature variations for the `17.7 K' and `18.8 K' resistance curves in above panel.
(c) \bpar magnetoresistance curves in same sample.} 
\label{RB-A}
\end{figure}

\figr{RB-A}(a) shows \rbb curves for various fixed $T$ for the \bper orientation ($B\perp$ film-plane) in SLCO sample A. There are pronounced oscillations over wide ranges of $T$ and $R$, superimposed on a steadily 
rising background MR 
(which generally follows the $R/R_n \sim B/B_{c2}$ flux-flow relation). 
The oscillations are not symmetric but have sharper minima in \rbby , which we will denote by \bxy . 
A graph of \bx vs count is very linear, indicating a high periodicity, and the slope of the straight-line fit yields a period of  $\Delta B_X = 0.149 \pm 0.004$ T independent of $T$. (A Fourier transform of \rbb produces the value $\Delta B_X = 0.154 \pm 0.008$ T.)  
The oscillations are strongest in the $B < 0.5$ T range and fade at higher $B$, although still well below the upper critical field \hcuy , with $R \ll R_n$, for the lower $T$; however, there is a hint of their reemergence at the top field. 
Comparing curves at different $T$, the oscillations get weaker and disappear as the normal state is approached, indicating that they are a feature related to the superconducting state and that the normal state MR is itself not oscillatory. In fact, the effect seems to be most prominent where the resistance responds most sensitively to the order parameter. 
The amplitude of the oscillations is so large (e.g., $\Delta R =  78 \ \Omega$, minimum to maximum, for the 17.7K curve) that it cannot possibly 
arise from a variation in $T$: the $dT/dR \approx 0.025$ K/$\Omega$ would imply a $\Delta T$= 1.9K, which is drastically higher than the 
$\pm 15$ mK maximum variation of the measured temperatures shown in panel (b). 

\figr{RB-A}(c) shows similar oscillatory \rbb curves for the \bpar orientation ($B \parallel$ film-plane), which have a comparable fractional amplitude superimposed on a less steep background MR (reflecting the much higher \hcu and lower flux mobility for \bpary). The periodicity of $\Delta B_X = 0.155 \pm 0.008$ T is the same as for \bper within error, indicating that the phenomenon has no connection with flux dynamics 
but has a deeper and more fundamental association with the \sg state. 

\begin{figure}[ht]
\includegraphics[width=0.95\hsize]{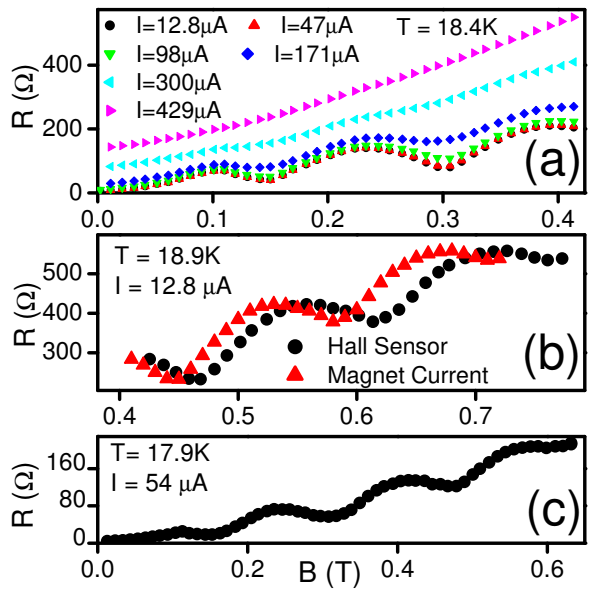}
\vspace{-1em} 
\caption{\em (a) Magnetoresistance curves at various dc currents. Oscillations are most prominent at low $I$, where the response is Ohmic, and disappear at higher $I$ (SLCO sample A in \bpery ).
(b) The magnetic field indicated by both the Hall sensor and electromagnet current produce comparable 
oscillations. (c) This curve was measured with a pulsed current and still shows oscillations.}
\label{OMR_vs_I}
\end{figure}

\figr{OMR_vs_I}(a) shows that the oscillations vanish as $I$ is increased, indicating that some delicate property of the \sg transition is needed; 
this action is similar to the effect of 
increasing $T$ seen in \figr{RB-A}(a). These observations establish that the oscillations are not an artifact of the voltage measurement, which is indifferent to the $I$, $B$, and $T$ of the sample. 
Could it be that there is something special happening in the apparatus around $T$\aby20 K and $R$\aby 100 $\Omega$, which serendipitously coincides with the transition in this \sscy ? No, because the oscillations have been seen in a variety of systems spanning the ranges $T$=4--74 K and $R$=0.01--1000 $\Omega$, with fractional resistance amplitudes ranging from zero to $>$50\%. 

Presence of weak links and microwave radiation can lead to Shapiro steps in the $I$-$V$ response\cite{shapiro}. \figr{OMR_vs_I}(a) shows the absence of these complications because the response is Ohmic for the lowest curves ($R(B)$ is independent of $V$ over a seven-fold range). Note that the entire range of effects related to weak links and noise, is incapable of producing \rbb oscillations with a $\Delta B_X$ independent of $T$ and geometry. 

\figr{OMR_vs_I}(b) shows \rbb curves with the abscissa taken in two ways: one uses the 
$B$ measured by the Hall sensor and the other is the $B$ estimated from the electromagnet current. Notwithstanding 
the small disagreement and offset in $B$ (expected from the hysteresis and non-linear response of the magnet's iron core), the oscillations are reproduced by both methods and are therefore not an artifact of the $B$ measurement. 
Furthermore the curves in \figr{OMR_vs_I}(b) were measured with a different  voltmeter (a Keithley 2000 multimeter instead of the 2182A nanovoltmeter) and were measured completely manually, to rule out artifact from a computer controlled data acquisition system. 

A final test, shown in \figr{OMR_vs_I}(c) demonstrates that the oscillations can also be seen in a pulsed transport measurement (involving an entirely different chain of electronics), if care is taken to minimize the introduction of spurious noise (careful isolation of the pulse generator and amplifiers from the mains power using isolation transformers and capacitors). In light of these extensive cross checks using multiple instruments (3 thermometers, 3 voltmeters, 3 current sources, and 2 $B$ measurements) 
we were not able to associate these oscillatory features to any artifact that could be produced by the apparatus. Therefore, as far as we can tell, it is an intrinsic phenomenon. 
  
\begin{figure}[ht]
\includegraphics[width=0.95\hsize]{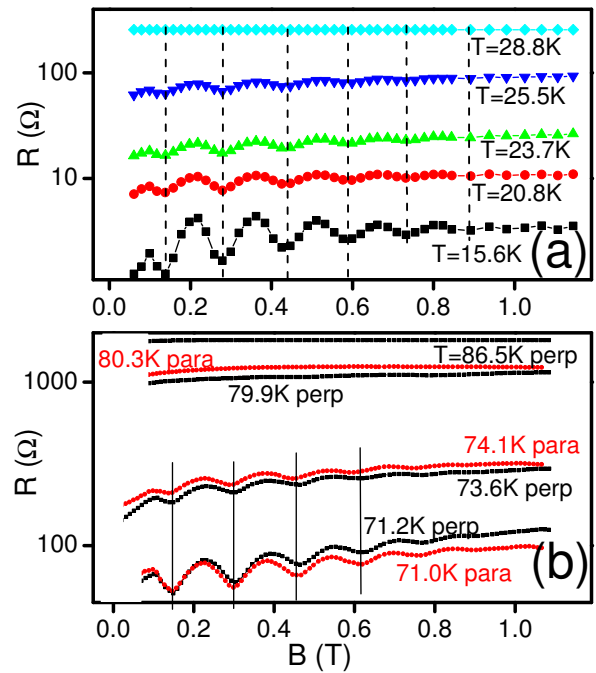}
\vspace{-1em} \caption{\em (a) \bpar oscillatory magnetoresistance in SLCO sample B. 
(b) \bpar and \bper oscillatory magnetoresistance in YBCO.}
\label{OtherSLCOandYBCO}
\end{figure}

\figr{OtherSLCOandYBCO} shows oscillatory \rbb curves for the other SLCO sample B (with very different dimensions) 
and the YBCO sample.
Note that YBCO is a hole doped cuprate superconductor unlike the electron-doped SLCO, and has a three times higher \tcy. Incredibly, these other samples have periodicities ($\Delta B_X=0.147 \pm 0.007$ T and $0.150 \pm 0.004$ T respectively) that are identical to the first SLCO sample within their error bars.
Additionally we have also observed the MR oscillations in the \ncco 
electron-doped cuprate superconductor and in the \bite  topological-insulator/iron-chalcogenide 
interfacial superconductor with similar $\Delta B_X$ periods (0.14 and 0.11 T)
despite very different material 
parameters and dimensions; those results will be described in detail elsewhere \cite{ncco-OMR}.
Not all \sg samples 
show the phenomenon. This unpredictibility of appearance may be expected from the anomalous states, as discussed below. 

So what is the physics that underlies our oscillations? 
Previously established oscillatory MR phenomena are incompatible with our observations. 
For example, multiply connected systems can have MR oscillations 
\cite{lpexpts}
because of thermally activated fluxoid dynamics \cite{fluxoid-dynamics}, 
current  excited vortices \cite{berdiyorov-wai}, or the Little-Parks \cite{lp} effect;
however, the samples would need to have a regular array of voids with a length scale of  $l \approx [\Phi_0/\Delta B_X]^{1/2}$=117 nm, 
and this non-existent arrangement would work only for a unique field orientation. 
Oscillatory MR occurs for $B$ close to \bpar due to 
rearrangements of the parallel vortex system \cite{brongersma}; however, 
the positions of the peaks depend on the field tilt angle. 
The Fulde-Ferrell-Larkin-Ovchinnikov (FFLO) modulated superconducting state \cite{fflo} could result in oscillations 
if its wavevector $q \sim 2\mu_B B/(\hbar  v_F)$ (in the microns range)
matches some intrinsic spatial periodicity, but this doesn't match any known length scale in our systems. Furthermore the present films do not have the special characteristics (e.g., clean, appropriate nesting of Fermi surfaces, etc.) that favor FFLO formation. 
The Shubnikov-de Haas effect \cite{ABR} is another scenario that produces MR oscillations; however, these are periodic in $1/B$ and occur at high $B$.
All of the above effects depend on the orientation of $B$, material parameters (Fermi-surface geometry, Fermi velocity, coherence length, etc.),  and/or sample dimensions and geometry. Thus none can explain our observed nearly universal periodicity.

We therefore turn to a fundamental $B$ dependent energy scale that occurs universally on a subatomic level:  the paramagnetic energy splitting $\Delta E_p=\pm \mu_B \Delta B_X\simeq 
9$ $\mu$eV (for our observed $\Delta B_X=0.15$T) between opposite spins; where 
$\mu_B=e\hbar/2m$  is the Bohr magneton. Cooper pairs ideally represent combinations of degenerate time-reversed states. Energy differences between these pair states tend to weaken the superconducting order parameter and increase $R$ (the familiar pair breaking concept). 
If the above mentioned paramagnetic shift were to cancel any preexisting energy differences between pair states with opposite spins, it would realign the pair energies 
and lead to a minimum in $R$. A candidate for such a spin-dependent energy splitting of comparable magnitude is the Lamb shift 
$\delta E_L$. The singular point like electron of the Dirac theory vibrates and appears spatially smeared (with a root mean squared displacement $\sqrt{\langle u^2 \rangle} \sim 10^{-13}$ m) because of 
its interactions with the zero-point electromagnetic oscillations of the vacuum. 
As a result its attraction is reduced for a singular potential well, such as the coulomb attraction of an atomic nucleus, 
and its energy is elevated. Low angular momentum states, especially $L$=0, that have a larger probability density 
at the potential's center undergo a larger shift and the magnitude of the shift depends on the depth  of the potential well (for example the Lamb shift for a given n and L is higher in high-Z atoms).  Thus different levels and different potential traps will undergo different shifts, leading to a sequence of minima, and we might expect the most elementary Lamb shift
$\delta E_L=mc^2(e^2/\hbar c)^5\simeq 11$ $\mu$eV 
(between the L=0 2S$_{1/2}$ and L=1  2P$_{1/2}$ doublet in a hydrogen atom) as an order of estimate of the basic energy scale, 
which is indeed comparable to $\Delta E_p$.
A possible candidate for the potential wells that nucleate the Lamb shift 
are the  anomalous states \cite{IVLEV} recently introduced to explain delayed x-ray laser bursts in 
metals \cite{xray}. These states are thin threads ($\sim 10^{-13}$ m), which can form randomly during sample preparation or through exposure to radiation. 
Because both $\Delta E_p$ and $\delta E_L$ are  effects arising on subatomic length scales, their magnitudes can be independent of the chemical nature and band structure of the material, and of the orientation of $B$, 
leading to a universal $\Delta B_X$ period as observed. (Since $\Delta E_p \sim \delta E_L \sim 1$ GHz, the effect would be especially sensitive to noise from digital wireless communications.) Since the effect is observed within broad transitions, weak links (which perhaps serve as preferred sites for the potential wells) may be relevant the mechanism. 

The above scenario of material-independent pair breakers is only suggested as one avenue for an explanation. We hope that our reported exerimental observations will stimulate the development of a more conclusive and rigorous theory. Given the striking nature of the oscillations, it may seem surprising that they have not been reported before. However, 
the effect is only observed in broad resistive transitions (most studies  tend to focus on ``high quality'' samples with sharp transitions), and the effect is 
sensitive to the level of excitation current and noise, and to fluctuations in $T$ (in our apparatus, special measures were taken to minimize these problems). If the $\Delta E_p \sim \delta E_L$ idea is 
on the right track, then this oscillatory effect could possibly be developed into a 
tool for 
investigating the anomalous states or other electronic level structure responsible for the phenomenon. 
Our preliminary 
investigations in magnetic superconductors show phase shifts and small reductions in $\Delta B_X$, possibly because of their internal magnetic fields. Thus our effect can potentially be used as a tool for measuring internal magnetic fields and permeabilities.

The authors would like to thank the following for providing samples, useful discussions, and other assistance: L. Fruchter, Z. Z. Li, 
M. Liang, J. M. Knight, N. S. Moghaddam, S. Varner, R. A. Webb, K. Stephenson, and N. Lu. This work was supported by the U. S. Department of Energy through grant number DE-FG02-99ER45763.

\end{document}